\documentclass{article}
\usepackage{natbib}
\usepackage{amsmath,amssymb}
\usepackage{xcolor}
\usepackage{graphicx} 
\usepackage{float}
\usepackage{pdflscape}
\usepackage{geometry}
\usepackage{amsthm, amssymb}
\usepackage[linesnumbered, ruled, vlined]{algorithm2e}
\usepackage{enumerate}
\usepackage{caption}
\usepackage{indentfirst}
\theoremstyle{definition}
\newtheorem{definition}{Definition}[section]

\theoremstyle{remark}

\title{RETRIEVING THE STRUCTURE OF PROBABILISTIC SEQUENCES FROM EEG DATA DURING THE GOALKEEPER GAME}
\author{Paulo R. Cabral-Passos, Priscila S. Azevedo,  Victor H. Moraes \\ Bia L. Ramalho, Aline Duarte, Claudia D. Vargas}
\date{July 2025}

\begin{document}

\maketitle

\begin{abstract}

    This work draws on the conjecture that fingerprints of stochastic event sequences can be retrieved from electroencephalographic data (EEG) recorded during a behavioral task. To test this, we used the Goalkeeper Game (game.numec.prp.usp.br). Acting as a goalkeeper, the participant predicted each kick in a probabilistic sequence while EEG activity was recorded. At each trial, driven by a context tree, the kicker chose one of three options: left, center, or right. The goalkeeper then predicted the next kick by pressing a button. Tree estimation was performed by applying the Context Algorithm to EEG segments locked to the button press (-300 to 0 ms). We calculated the distance between the penalty taker’s tree and the trees retrieved per participant and electrode. This metric was then correlated with the goalkeeper's success rates. We observed a clear reduction in the overall distance distribution over time for a subset of electrodes, indicating that EEG dependencies become more congruent with the penalty taker's tree as the goalkeeper learns the sequence. This distance is inversely proportional to the goalkeepers’ success rates, indicating a clear relationship between performance and the neural signatures associated with the sequence structure. 
    
\end{abstract}

\textbf{key-words}: statistical learning; temporal sequences; electroencephalography; structural learning.

\section{Introduction}

The present study addresses the statistical brain conjecture, originally proposed by \citet{Helmholtz1867}, which states that the brain automatically infers upcoming events based on previous experience. In its contemporary version, this conjecture suggests that fingerprints of sequences of events are encoded within the brain’s activity and can be recovered using appropriate analytical methods \citep{Duarte2019, Hernandez2021, CabralPassos2024, Hernandez2024, Dehaene2015, Planton2021, AlRoumi2023}. These studies build upon the principle of Minimal Description Length (MDL) developed by Rissanen (1978). This principle states that the model that best explains the data is the one that can be mapped to its shortest description. This fundamental approach motivated the development of new models for stochastic sequences with important applications in neuroscience \citep{Duarte2019, Hernandez2021, CabralPassos2024, Hernandez2024, Dehaene2015, Planton2021, AlRoumi2023}. 

In 1983, \citet{Rissanen1983} highlighted that in many real-life sequences, the next event identification often depends on a suffix, or a subsequence of the recent past events. Furthermore, upon each new event observation, the length of this suffix can vary \citep{Rissanen1983}. Rissanen used the term context to refer to the shortest suffix that contains all the available information necessary to predict the next event. Although a sequence can be infinitely long, its set of contexts is often finite. Thus, knowing all contexts within a sequence is equivalent to possessing its structural information, regardless of its size. These contexts can be represented as a model in the form of an upside-down rooted tree, where the branches lead to labeled leaves corresponding to the sequence contexts. Consequently,  the sequence can be considered as a realization of the associated context tree model, encapsulating its underlying structure.

The framework developed by \citet{Duarte2019} draws inspiration from \citep{Rissanen1983, Rissanen1983b}. It proposes a new method to identify the structure of stimulus sequences in brain activity. To apply this method, a participant is exposed to a sequence of auditory stimuli. These stimuli follow a probabilistic context tree model, also known as a Variable Length Markov Chain. During this procedure, the participant's brain activity is recorded using electroencephalography (EEG). The time series of EEG activity is then segmented and systematically paired with each stimulus. With both the sequence of stimuli and the associated EEG segments, one can model their relationship. This makes it possible to verify whether they share the same past dependencies.

In \citet{Hernandez2021}, the method proposed in \citet{Duarte2019} was tested. By combining the Context Algorithm \citep{Duarte2019} and the Projective Method \citep{CuestaAlbertos2007}, it was possible to model the relationship between the sequence of stimuli and the sequence of associated EEG segments. This procedure allowed for the retrieval of context trees from the EEG signal. When the resulting context tree matches the one driving the auditory stimuli, this indicates that their past dependencies are compatible. Employing this approach, \citet{Hernandez2021} demonstrated that the context trees derived from the EEG signals collected from electrodes placed above the frontoparietal regions of the brain matched the context tree driving the auditory stimuli at the group level.
 
Employing a similar framework, \citet{CabralPassos2024} used the Goalkeeper Game to deliver game events. Developed by NeuroMat/FAPESP, the Goalkeeper Game is a computer game that simulates a real-life scenario where learning the sequence's structure is crucial to achieving the task. Once the structure of the penalty taker's choice is identified, players can predict the direction of the next kick. The penalty taker's choices follow a probabilistic context tree model. Participants act as goalkeepers and are instructed to avoid as many goals as possible. The purpose of the study was to model the relationship between penalty takers' choices and goalkeepers' response times. \citet{CabralPassos2024} showed that the context trees obtained from modeling the goalkeeper's response times corresponded to the context tree generated by the penalty taker's choices at the group level.

The present study follows directly \citet{CabralPassos2024}. We utilize the framework developed in\citep{Duarte2019,Hernandez2021,Hernandez2024,CabralPassos2024} to quantitatively assess the relationship between sequence-related EEG measurements and behavioral data. We asked participants to play the Goalkeeper Game while monitoring their brain activity. Acting as goalkeepers, the participants should try to predict how the sequence of the penalty taker unfolds. We correlated the success rate of the goalkeepers with the context tree derived from the EEG segments in the vicinity of their motor responses. Our findings indicate that the distance between the context tree of the penalty taker and the one estimated from the goalkeepers' EEG is inversely proportional to the goalkeeper´s success rates.

\section{Methodology}

\subsection{Experimental protocol}

This experiment was conducted in compliance with all relevant guidelines and protocols, with approval from the Ethics Committee of the University of São Paulo (CAAE 69431623.2.0000.5407) and the Deolindo Couto Institute of Neurology at the Federal University of Rio de Janeiro (CAAE 72756723.8.-0000.5261). All participants provided informed consent. Identical experimental set-ups were used at two locations: the Research, Innovation and Dissemination Center for NeuroMathematics (NeuroMat/FAPESP) at the University of São Paulo and the Neurosciences and Rehabilitation Laboratory at the Federal University of Rio de Janeiro. The study included 26 right-handed participants (12 female, mean age 24 ± 6.61 years), whose handedness was confirmed using the Portuguese version of the Edinburgh Handedness Inventory.

Participants played the Goalkeeper Game (game.numec.prp.usp.br), an electronic game developed by NeuroMat, where they acted as goalkeepers during a series of penalty trials. Each participant was seated comfortably with arm and foot rests, and their nasion was aligned with the vertical midpoint of a 32-inch Display++ monitor (Cambridge Research Systems Ltda.) positioned 114 cm away. Response times and EEG signals from 32 electrodes were recorded throughout the game.

\begin{figure}[H]
    \centering
    \includegraphics[width=0.95\linewidth]{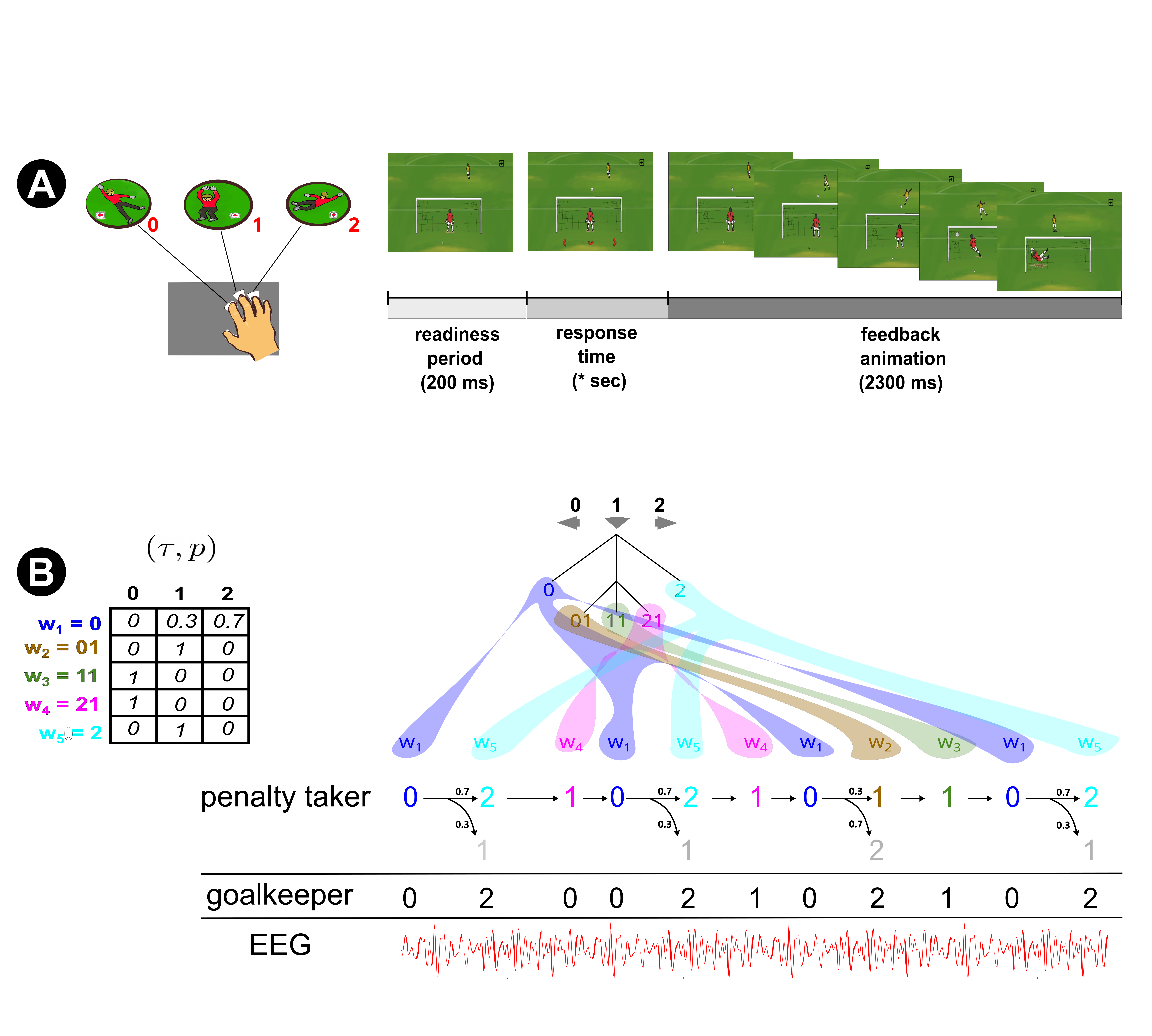}
   \caption{The Goalkeeper Game task. A) Key/finger/choice mapping and trial configuration: on the left, we show the keyboard design and the fingers' correspondence with the game option: right index finger (0), right middle finger (1), and right ring finger (2). On the right, the duration of each trial section is presented along with the game screen samples. The first section is termed the readiness period, during which the goalkeeper and penalty taker on the screen remain at rest. The second is termed the response time section, marked by the appearance of three arrows at the bottom of the screen informing the goalkeeper to inform his/her prediction. No maximum duration for this period was set. This section ended with the button press, which triggers the feedback animation ($2300ms$). B) Context tree model generating the penalty taker's choices and examples of penalty taker's and goalkeeper's sequences. The bottom trace represents the continuous recording of the EEG signal. The arrows indicate the transition probability following the context tree model; the transition that did not occur is presented in gray.}
    \label{fig:task}
\end{figure}

The game version implemented 1500 trials. In each trial, the goalkeeper should predict the penalty taker's next choice by pressing a button on a special keyboard (CEDRUS RB-840). The penalty taker might shoot to one of three options: to the left, to the center, or to the right of the goal, henceforth referred to as 0, 1, and 2, respectively. If the choices of both the penalty taker and the goalkeeper coincided, an animation at the end of the trial would show a successful defense; otherwise, a goal animation was presented. Figure \ref{fig:task}A shows a schematic representation of the trial. On the left, we show the keyboard design and the fingers' correspondence with the game option: right index finger (0), right middle finger (1), and right ring finger (2). On the right, the duration of each trial section is presented along with the game screen samples. The first section is the readiness period, during which the goalkeeper and penalty taker on the screen remain at rest ($200ms$). The second section is the response time section, marked by the appearance of three arrows at the bottom of the screen, informing that the goalkeeper is allowed to inform his/her prediction. No maximum duration for this period was set. The response time section ended with a button press, immediately initiating the feedback animation ($2300ms$). There were no rest intervals between trials, only two pauses, each after 500 trials. During this period, the participant was allowed to move, drink water, and communicate any concern to the researcher, notifying the researcher when he/she felt ready to return to the game.

Before starting the game, the goalkeepers were presented to a special screen with information about which fingers he/she should use, the finger/button/choice mapping, and the game mechanics. The following instruction was presented:  "Your task is to avoid as many goals as you can. Attention: the penalty taker is not influenced by your choices". The penalty taker's choices followed the probabilistic context tree model represented in Figure \ref{fig:task}B. On top of the table, the symbols 0, 1, and 2 represent the penalty taker choices. Each table row presents a context on the far left. The model presents a total of five contexts. At any step of the sequence, the next symbol of the sequence can be predicted by identifying in which context the step ends. For a given context, the values under each penalty taker choice (0,1,2) indicate the probability associated with this symbol being the symbol immediately after the context. As an example, if the current step ends with the context 01 (that is, left, center), the next choice of the penalty taker will always be 1 (center). On the other hand, if the current step ends in the context 0, the next choice can be either 1 (center) with probability of $0.3$ or 2 (right) with probability of $0.7$. Therefore, whenever this context appears in the penalty taker sequence, it is impossible to predict with certainty the next choice. A context is described as the minimum amount of past symbols necessary to predict the next symbol. The $\tau$ symbol on top of the table indicates the set of contexts $\tau = \{ 0, 01, 11, 21, 2 \}$, while $p$ indicates the probabilities associated with each context. On the left, a penalty taker sequence generated by the model is presented as an example; the colored pathways linked to the upside-down tree representation of $\tau$ indicate the context associated with each choice of the penalty taker. Right at the bottom, we present an example of a sequence of predictions of the goalkeeper.

\subsection{EEG data processing}

EEG data were recorded at a sampling rate of 2500 Hz using 32 electrodes (ActicHamp, Brain Products GmbH), positioned according to the International 10-20 System. A conductive paste was applied to keep the electrode impedance below 25 $k \Omega$. During acquisition, the Cz electrode served as the reference. For preprocessing, the data were first re-referenced to the average of all electrodes and then high-pass filtered using a zero-phase 4th-order Butterworth filter with a cutoff frequency of 1 Hz. Independent Component Analysis (ICA) was then applied, followed by classification using ICLabel. Components with a probability of less than $10 \%$ of representing brain activity were removed. Finally, the data were low-pass filtered with a zero-phase 4th-order Butterworth filter with a 45 Hz cutoff and then downsampled to 256 Hz.

\subsection{Linear Mixed Effect Model}

A linear regression model can be described as follows. Consider $Y$ to be a dependent random variable on which we intend to perform a prediction. Therefore, we can model its relationship with an independent variable $x$ as
\begin{equation}\label{eq:LinReg}
Y = \beta_{0} +\beta_{1} x + U,
\end{equation}
where $\beta_{0}$ and $\beta_{1}$ are fixed effects and $U \sim \mathcal{N}(0,\sigma)$ is a random effect due to measurement error. In a Linear Mixed Effect Model (LMEM)\citep{Henderson1953}, we consider other random effects related to each group. Indeed,
let $M$ be the number of groups in which \eqref{eq:LinReg} is observed between two variables of interest. We let $j = 1, \ldots, M$ be an index that identifies each group. The total number of samples in our LMEM is
\[
N = \sum_{j=1}^{M} n_{j},
\]
where $n_{j}$ is the number of samples in the group $j$. The model itself can be written as
\[
\vec{Y} = \mathbf{X}\vec{\beta}+\mathbf{Z}\vec{r}+\vec{U}
\]
where $\vec{Y}$ is a vector containing the $N$ observations of our dependent random variable, $\mathbf{X}$ is a matrix whose first column is a vector of ones that, along with the unknown fixed intercept in the first row of $\vec{\beta}$, is equivalent to $\beta_{0}$ in the simple linear regression. In a model with a single dependent variable, such as the one in our analysis, $\mathbf{X}$ has only one additional column, which contains the observations of the dependent variable. $\mathbf{Z}$ is a $N\times M$ sparse vector in which the $j$-column contains rows of ones in the observations related to group $j$ and zero otherwise. $\vec{r}$ is a vector of $M$ random variables that composes the random counterpart of the fixed effects in  $\vec{\beta}$. In fact, each row is an i.i.d. random variable with distribution $\mathcal{N}(0,\sigma_{z}), z>0$. Finally, $\vec{U}$ is a vector of size $N$ in which each row contains an i.i.d. random variable with distribution $\mathcal{N}(0,\sigma)$, which accounts for the measurement error. For more details on other aspects of the LMEM, we refer the reader to \citet{Wikistat2016}. 

For each window and electrode, centroids were calculated by taking the median of the distances for all participants; we will henceforth refer to them as median-based distances. These measures were used to predict the corresponding median-based success rates using the LMEM. A sliding window method was used to obtain each window of analysis. Each window corresponded to 300 trials, and two subsequent windows presented a superposition of 50 trials. Therefore, we can retrieve the first and last trial for a given window $i = 1, \ldots, T$, where $T$ is the total number of windows, by applying the following rule: $initial~trial = 1 + (i-1)50$ ;  $last~trial = 300 + 50(i-1)$. Considering the last trial of the first window, we can calculate how many windows are necessary for no superposition between two windows. Solving $1 + (i-1)50 > 300$ for $i$, we get that the superposition ends every six windows. In the LMEM, we defined the groups for each window to avoid superposition. This way, the first group, for example, was composed of the 1st, 7th, 13th, 19th, and 25th windows.

\subsection{Basic mathematical notations and definitions}

Let $A$ be a finite set, here called \textit{alphabet}. Given $k \geq 0$, we denote as $A^{k}$ the set of all sequences of $k$ symbols of $A$ and $A^{*} = \cup_{k = 1}^{\infty} A^{k}$ the set of all possible finite sequences of $A$.
For any $m,n \in \mathbb{Z}$ with $m \leq n$, $v_{m}^{n}$ denotes the sequence $(v_m,v_{m+1}, \ldots, v_{n})\in A^{n-m+1}$ of length $l(v_{m}^{n}) = n - m + 1$ of symbols of A. 

A sequence $u\in A^{k},$ with $1\leq k\leq n-m+1,$ is said to be a subsequence of $v=v^n_m\in A^{n-m+1}$ if there exists $ m\leq j \leq n-k+1$ such that $u = v_j^{j+k-1}$. Given two sequences $u=u^n_m\in A^{n-m+1}$ and $v=v^s_r\in A^{s-r+1}$, their \textit{concatenation} is the sequence  $uv = (u_m,\ldots,u_n,v_r\ldots,v_s)\in A^{n-m+s-r+2}$ whose length is $l(u) + l(v)$ and both $u$ and $v$ are subsequences of $uv$.

The sequence $s$ is said to be a \textit{suffix} of $v$, indicated as $ s \preceq v$, if there exists a sequence $u$ such that $v = u s$. For the specific case in which $v \neq u$, indicated as $s \prec v$, we say that $s$ is a \textit{proper suffix} of $v$. For any finite sequence $u=u_{m}^{n}$, we denote by $suf(u)$ the immediate proper suffix of $u$ given by $u_{m+1}^{n}$, with the convention that $suf(u)=\emptyset$ when $m=n$.

\begin{definition}[context tree]
A finite subset $\tau$ of $A^{*}$ is a \textit{context tree} if it satisfies the following properties.
\end{definition}
    \begin{enumerate}[i]
    \item \textit{Suffix property}: there are no two sequences $u,v \in \tau$, such that $u \prec v$ or $v \prec u$.
    \item \textit{Irredutibility}: replacing any sequence $v \in \tau$ by a suffix  $ s \prec v$ implies the violation of $i$.
    \end{enumerate}

In the present work, we are interested in the alphabet $A = \{ 0, 1, 2 \}$ which represents the goalkeeper's and penalty taker's choices, that is, left (0), center (1), and right (2). 

\subsection{Context algorithm}

In the algorithm description that follows, $X_{1}^{N}$ represents a sequence of length $N$, corresponding to the choices made by the penalty taker. For any given subsequence $w$ of $X_{1}^{N}$, the EEG segment associated with the $j$-th occurrence of $w$ is denoted as $y_{(w,j)}$ and belongs to the set of squared-integrable functions $L^{2}(\mathbb{R}^{d})$ of length $d$. Correspondingly, $\mathcal{Y}_{w}$ represents the set of all EEG segments associated with $w$ in $X_{1}^{N}$ and $\mathcal{Y}^{*}$ the set of all EEG segments associated with any subsequence of $X_{1}^{N}$.

\begin{algorithm}[H]
\KwIn{ $X_{1}^{N}$, $\mathcal{Y}^{*}$, $k$ } 

$\hat{\tau} \gets \varnothing$ \\
$C \gets \{ u \in A^{k} : u\text{ is a subsequence of } X \}$ \\
\While{ $C \neq \varnothing$}{
    $w \leftarrow \text{ element sampled from C}$ \\
    $s \leftarrow suf(w)$ \\
    $F(w) \leftarrow \{ v = a s : a \in A, v \in C \}$ \\
    \eIf{ $F(w) \subset C$}{
        \eIf{ $F(w)$ has a single element}{
            $C \leftarrow C \setminus F(w)$ \\
            $C \leftarrow C \cup \{ s \}$
        }{
            \eIf{ $Decision$: $F(w), \{ \mathcal{Y}_{v}: v \in F(w) \} \rightarrow 1 $}{
                $\hat{\tau} \leftarrow \hat{\tau} \cup F(w)$ 
            }{
                $C \leftarrow C \cup \{ s \}$
            }
            $C \leftarrow C \setminus F(w)$
        }
    }{
        $\hat{\tau} \leftarrow \hat{\tau} \cup (F(w) \cap C )$ \\
        $C \leftarrow C \setminus (F(w) \cap C )$
    }
}

\KwOut{ $\hat{ \tau}$}
\caption{Context Algorithm}
\end{algorithm}

The decision in line $12$ is done by testing all possible pairs in $\{ \mathcal{Y}_{v}: v \in F(w) \}$. For each pair $(\mathcal{Y}_{v},\mathcal{Y}_{v'})$ the projective method\citep{CuestaAlberto2021} is used to test the null hypothesis that the EEG segments from $\mathcal{Y}_{v}$ and $\mathcal{Y}_{v'}$ are generated by the same law as in \citet{Hernandez2021}. The algorithm is illustrated in Figure \ref{fig:R1}A-D.

\subsection{Balding distance ($d$)}

Following \citet{Balding2009}, we will use a distance between trees herein denoted as $d$. Consider $\mathcal{T}_{ \tau } = \cup_{w \in \tau} \{ u: u \preceq w  \}$ and $\mathcal{T}_{ \tau' } = \cup_{w \in \tau'} \{ u: u \preceq w  \} $. We define $d$ as
$$
d( \tau , \tau' ) = \sum_{ u \in \mathcal{T}_{ \tau } \cup \mathcal{T}_{ \tau' } } | \textbf{1}_{ \{ u \in \mathcal{T}_{ \tau } \} } - \textbf{1}_{ \{ u \in \mathcal{T}_{ \tau' } \} } |z^{l(u)}
$$
\noindent where $0 < z < 3^{3/2}$. The parameter $z$ was set to $0.5$ in the following analysis.

\subsection{Success rates}

Let the $X_{m}^{n}$ and $Y_{m}^{n}$  be the sequences of penalty taker's and goalkeeper's choices across the trial $m$ to $n$, respectively. Then we define the success rate across these trials as
$$
    S(X_{m}^{n},Y_{m}^{n}) = \frac{\sum_{k = m}^{n} \textbf{1}_{ \{ X_{k} = Y_{k}\} }}{n-m+1}
$$

\begin{figure}[H]
    \centering
    \includegraphics[width=1\linewidth]{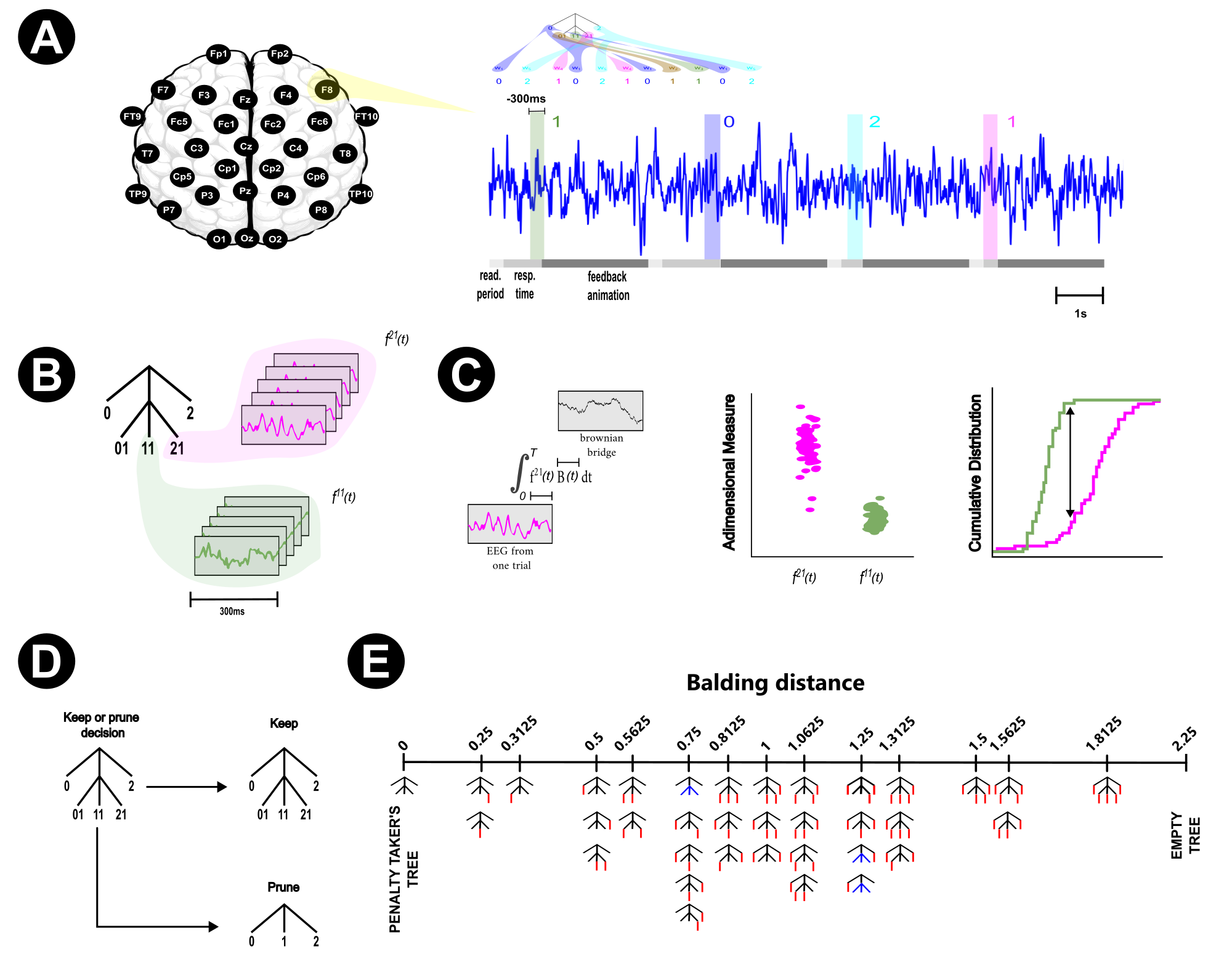}
    \caption{\textbf{Context Algorithm and distance calculation.} A) The cartoon on the left shows the spatial distribution of the 32 electrodes following the 10-20 International System from which the EEG signals were recorded. On the right, we show an example of an EEG recorded from a single electrode of one participant. The gray bars on the bottom indicate the readiness period, response time, and feedback sections. The colored windows indicate the 300 ms of the EEG signal of each trial used for the context tree estimation. The color code indicates the context of each trial following the tree representation (top left corner). B) The pruning procedure was performed by (1) selecting not unvisited terminal subtrees of the complete tree. In our example, the branch corresponding to symbol 1. (C) testing the null hypothesis of equality of distributions for the sample of EEG segments associated to the leaves of that subtree,  and (D) pruning the subtree if the null hypothesis was not rejected for all pairs of leaves in the subtree or keeping the subtree if the null hypothesis was rejected for at least one pair of leaves. E) The goalkeeper’s estimated context tree was then compared with the penalty taker’s tree using the distance as in \citet{Balding2009}. The scale shows the possible distances and the corresponding trees in a simplified representation. The red lines indicate branches that should have been pruned during the context tree estimation, and the blue lines indicate branches that should have been spared. 
}
    \label{fig:R1}
\end{figure}

\break

\section{Results}

\indent The Context Algorithm was used to model the relationship between the penalty taker’s sequence and the corresponding sequence of EEG segments recorded from each goalkeeper’s electrode within a given window of analysis. These EEG segments corresponded to the 300ms preceding the button press that triggered the feedback animation (see Figure \ref{fig:R1}A).

To ensure that this time window was indeed capturing the brain activity associated with the goalkeepers’ motor preparation, event-related potentials were estimated considering the interval between the end of the feedback of the previous trial and the next button press (see Supplementary Methods session). 

\begin{figure}[H]
    \centering
    \includegraphics[width=0.9\linewidth]{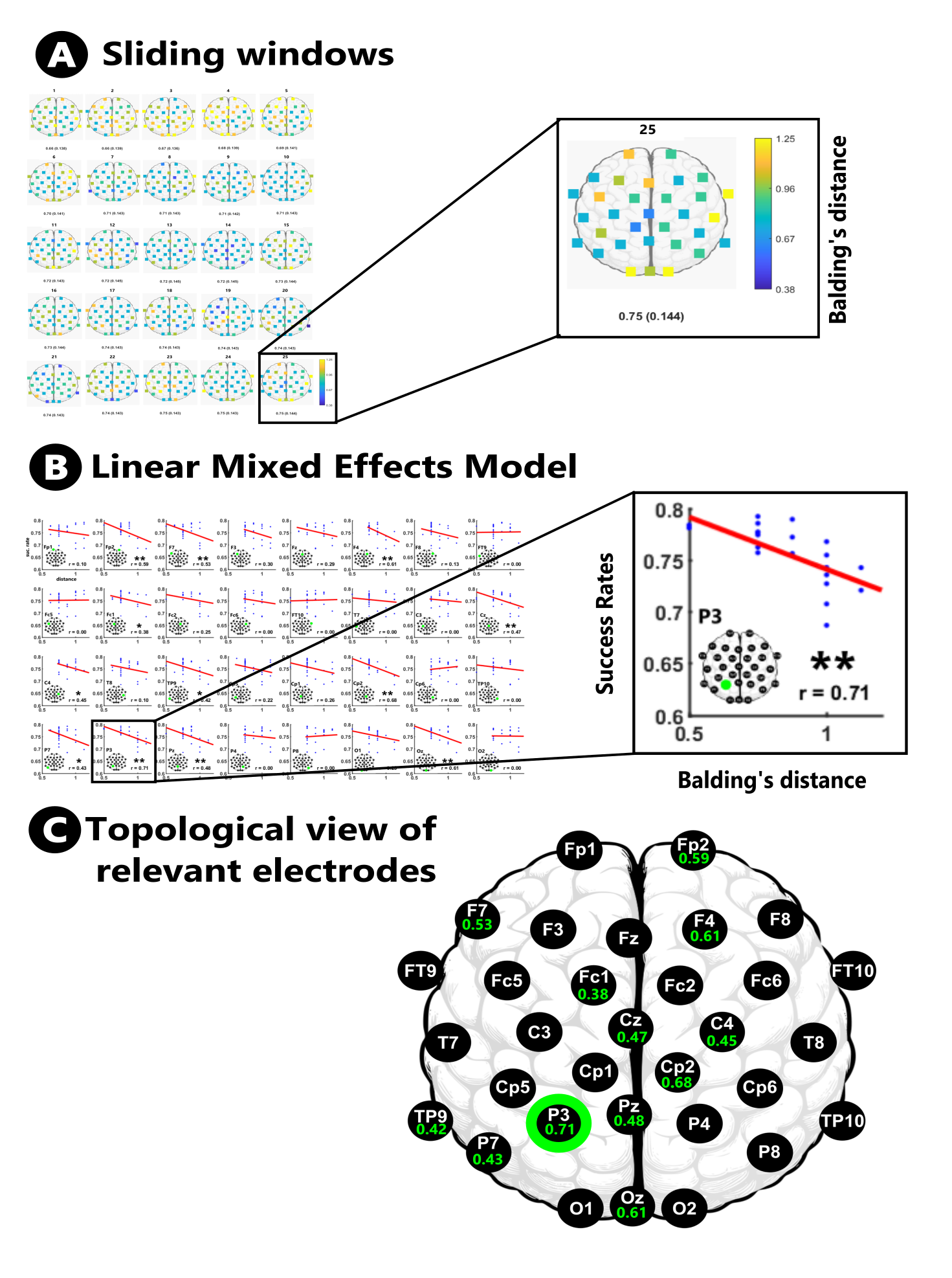}
    \caption{\textbf{Processing Pipeline} A) \textit{Sliding Windows}, context trees were estimated for 300 trial windows with 50 trials of displacement. Then, the distances between the penalty taker’s tree and each goalkeeper’s estimated tree were calculated per window and electrode. Top numbers indicate the window’s index. The colorbar shows the distance-color mapping. Mean and standard deviation of the success rate are presented at the bottom of each topo-plot. The time evolution suggests an inverse relationship between the distances and the success rates. B) \textit{Linear Mixed Effects Model}, success rate as a function of Balding’s distance. The model was used to evaluate the predictability of the success rate based on  Balding’s distance. Each graph shows a single electrode highlighted in green. Each blue dot represents a median-based centroid of the distances paired with the median-based success rate centroid for one window (* $p<0.05$ and ** $p<0.01$). The absolute values of the correlations (r) are presented for each graph. The inverse relationship for specific electrodes indicates that the corresponding brain regions are enrolled in the structural learning of the penalty taker's sequence. C) \textit{Topological view of relevant electrodes} Electrodes that present a significant correlation in the relationship between success rate and Balding's distance.}
    \label{fig:RG}
\end{figure}

The estimated context trees were then compared with the context tree used to generate the penalty taker’s sequence of choices employing Balding’s distance (Balding et al., 2009). This distance reflects how similar the past dependencies of the penalty taker’s and the goalkeeper’s context tree are. Figure \ref{fig:RG}A shows the time evolution of Balding’s median distance for each electrode across the 25 windows. The 25th window was highlighted for illustrative purposes (see Figure S3 for details of each window).  Blue indicates distances closer to zero (with a value of zero indicating identical trees) while yellow corresponds to distances far from zero, increasing as the trees become less similar. The number on top of each brain scheme indicates the window’s index, and the values on the bottom correspond to the mean success rate and the corresponding standard deviation for the whole set of participants within that window. For several electrodes, we see a transition from yellow to blue as the window’s index increases. Likewise, an increase in the mean success rates across windows can be verified, suggesting an inverse relationship between the distances and success rates.

A Linear Mixed Effects (LME) model was used to investigate the existence of an inverse relationship between success rates and Balding’s distance retrieved from the estimated trees. For each window and electrode, the median-centroid distance was calculated for the trees estimated from the 26 participants, where the median-based centroid distances were taken as the independent variable, while the median-based centroid success rates were taken as the dependent variable in the model. Figure \ref{fig:RG}B shows the success rates as a function of Balding’s distance for the set of 32 electrodes. Results of P3 are highlighted for illustrative purposes (see Figure S4 for details of each electrode). Each electrode is identified by its corresponding label in the 10-20 system, and its position is highlighted on the brain top-plot by a green solid circle. For each electrode, blue dots correspond to the median-based centroid of the distance paired with the median-based centroid of the success rate for a single window. A detailed summary of the model parameters and associated statistics can be found in Supplementary Table TS1. Correlation values are shown on the right side of each brain top-plot. Significant correlations were found for the following electrodes (Fp2, F7, F4, Fc1, Cz, C4, Cp2, TP9,  Pz, P3, P7, Oz) and  ranged from $38\%$ at the left frontocentral region (Fc1) to $71\%$ at the left parietal region (P3). Figure \ref{fig:RG}C shows the correlation values only for the electrodes with significant results and the spatial distribution of the electrodes. Larger correlation values can be observed in frontal regions of the right hemisphere (F4 and Fp2) and in the parietal regions (P3) of left hemisphere. Along the midline, the larger correlation value is verified over the occipital region (Oz). 

\section{Discussion}

In this study, sequences of events were generated by combining external events with participants’ stimulus-related predictions within a sensory-motor learning paradigm. Participants played the Goalkeeper Game, in which they took on the role of a goalkeeper facing a sequence of 1500 penalty kicks, attempting to predict each penalty taker's decisions. We applied a mathematical framework to analyze the relationship between the past dependencies of these events and those of the EEG signal segments recorded just before each motor response detection. Both the penalty taker’s decision algorithm and the goalkeeper’s predictive model, which was derived from the EEG activity, can be represented as context trees, enabling a direct comparison. This comparison was done by employing a tree-space distance metric, where a distance of zero indicates identical trees and larger values reflect increasing dissimilarity. Our results show that, for frontal, central, temporoparietal, centroparietal, parietal, and occipital regions, this distance is inversely proportional to the goalkeepers’ success rates, indicating a direct correlation between the success rates and the neural signature related to the structure of the sequence of events.

Our results show that, for the EEG electrodes with significant results, poorer prediction performances are associated with task-simultaneous EEG activity that statistically deviates from the past dependencies of the penalty taker sequence. To verify this, we recorded participants’ EEG while they were presented with a sequence composed of fixed and non-fixed context-dependent transitions. In \citet{Bogaerts2020}, a similar approach was taken to investigate the relationship between learning and brain activity, using sequences that also included fixed and non-fixed transitions. Their protocol was asynchronous, meaning that participants were first exposed to the sequence structure and then subjected to a two-alternative forced choice task, where they had to distinguish between valid and invalid sequence snippets. Their findings revealed that the amplitude difference in the EEG beta band between fixed and non-fixed transitions, measured over a specific cluster of electrodes, showed a positive correlation with participants’ test performance. While these results align with our findings, a few important things are worth mentioning regarding the pros and cons of it. Choosing an EEG frequency band for analysis imposes a predefined dimensionality reduction that, while it may yield a better signal-to-noise ratio, can also discard information outside the spectral window (for a discussion see \citet{Widmann2015}). Our approach is chosen so as to define the dimensionality reduction later, based on testing whether the EEG associated with two strings of the sequence differ in their statistical structure, leaving room for the identification of differences anywhere in the frequency spectrum. This can be a better option given it is known that beta, alpha \citep{Monroy2019} and even gamma \citep{Saringer2023} oscillations have also been involved in statistical learning when visual information plays a role in the stimulus sequence, which is the case both here and in \citet{Bogaerts2020}. Furthermore, although not within the scope of this study, our approach allows objective investigation of differences in sequences sharing the same alphabet and with similar EEG–EEG-performance correlations as can appear in protocols such as \citet{Hernandez2021}. This may reveal behavioral properties shared by different sequences, associated neural signatures, and help refine hypotheses of future studies.

Regarding the topology of our results, a significant association was found between participants’ performance and EEG signals for several regions of the brain, notably the frontal, central, centro-parietal, parietal, and occipital areas ($p < 0.01$). These findings are partially in agreement with those reported by \citet{Wang2017}, who used fMRI to track cortico-striatal activity during visual statistical learning tasks. According to their study, the cortical regions involved included the prefrontal cortex, sensory-motor areas, and occipito-temporal regions. Although we did not find associations with the occipito-temporal areas, our EEG data, mapped to their respective anatomical locations, identified Fp2 (prefrontal cortex), Cz, and C4 (sensory-motor areas) as performance-related. However, some electrodes showing the strongest correlations with performance, namely F4, Cp2, P3, and Oz, do not correspond directly to the regions highlighted in \citet{Wang2017} study. This discrepancy may be due to methodological differences: their analysis focused on brain activation during the task via fMRI, while our study relates brain activity to performance outcomes using EEG. Regarding the topological characteristics associated with the signal-performance relationship in \citet{Bogaerts2020}, the comparison ends up being more difficult, since the set of electrodes from which the relationship is extracted varies across the protocol sections, without a strong pattern emerging.

We focused our analysis on movement preparation due to established associations between statistical learning and both (1) behavioral measures, such as response and reaction times \citep{CabralPassos2024, Nissen1987, Hunt2001, Frost2015, Wang2017, Notebaert2009}, and (2) corticospinal excitability \citep{Bestamann2008}. A well-characterized marker of movement preparation in EEG is the Contingent Negative Variation (CNV) \citep{Sznabel2023}. The ERP technique used for CNV extraction allowed us to identify and validate the movement preparation window in our analysis, with protocol-specific adaptations to account for the absence of a maximum time limit for participant responses (see Supplementary Material; methodological details in Figure S1). As shown in Supplementary Figure S2, the CNV emerged predominantly during the final 300 ms preceding participants' responses and was most pronounced over midline central and frontal electrodes. In particular, although the CNV reliably indexed motor preparation temporally, despite sharing the same time window with our analysis, its spatial distribution did not fully overlap with the scalp sites where the EEG signal was shown to be more strongly correlated with the behavioral performance of participants.

The present study is based on the principle of Minimum Description Length (MDL) developed by Rissanen \citep{Rissanen1983, Rissanen1983b}. This principle states that the model that best explains the data is the one that can be mapped to its shortest description. It motivated the development of models for stochastic sequences that inevitably found their way in Neuroscience \citep{Duarte2019, Hernandez2021, CabralPassos2024, Hernandez2024, Dehaene2015, Planton2021, AlRoumi2023}. The theory that the brain uses this MDL principle in order to store data is highly plausible and largely accepted. From these models, the one developed by Rissanen \citep{Rissanen1983} himself, named Context Tree Models or Variable Length Markov Chains, allows the unification of both deterministic and non-deterministic patterns arising in sequence data in a single model that ultimately follows the MDL. A rigorous mathematical framework has been built for this theory \citep{Galves2008,Galves2012,Galves2013,Garivier2011}, allowing us to finally test this hypothesis. Here, we present strong evidence showing how behavior and neural activity can be tied by the MDL in statistical learning, paving the way for further investigation.

\section{Acknowledgment}
S.Paulo Research Foundation (FAPESP)grants: 2022/00699-3, Research, Innovation and Dissemination Center for Neuromathematics (2013/ 07699-0) FAPERJ (grant CNE 202.785/2018; grant E-26/010.002418/2019;  CNE E-26/204.076/2024 and $\#$E-26/200.349/2025 to BLR), FINEP (grant 18.569-8/2011), CNPq (grant 310397/2021-9; grant 407092/2023-4) and CAPES (grant 88887.671450/2022-00). The author's thank Italo Ivo Lima Dias Pinto for reviewing the figures in the manuscript.

\bibliographystyle{unsrtnat}
\bibliography{ref}

\begin{thebibliography}{33}
\providecommand{\natexlab}[1]{#1}
\providecommand{\url}[1]{\texttt{#1}}
\expandafter\ifx\csname urlstyle\endcsname\relax
  \providecommand{\doi}[1]{doi: #1}\else
  \providecommand{\doi}{doi: \begingroup \urlstyle{rm}\Url}\fi

\bibitem[von Helmholtz(1867)]{Helmholtz1867}
Hermann von Helmholtz.
\newblock \emph{Handbuch der physiologischen Optik}, volume III.
\newblock Leopold Voss, Leipzig, 1867.
\newblock Translated by The Optical Society of America in 1924 from the third German edition, 1910, as \textit{Treatise on Physiological Optics}.

\bibitem[Duarte et~al.(2019)Duarte, Fraiman, Galves, Ost, and Vargas]{Duarte2019}
A.~Duarte, R.~Fraiman, A.~Galves, G.~Ost, and C.~D. Vargas.
\newblock Retrieving a context tree from eeg data.
\newblock \emph{Mathematics}, 7\penalty0 (9), 2019.
\newblock \doi{https://doi.org/10.1038/s41598-021-83119-x}.

\bibitem[Hernández et~al.(2021)Hernández, Duarte, Ost, Fraiman, Galves, and Vargas]{Hernandez2021}
N.~Hernández, A.~Duarte, G.~Ost, R.~Fraiman, A.~Galves, and C.~D. Vargas.
\newblock Retrieving the structure of probabilistic sequences of auditory stimuli from eeg data.
\newblock \emph{Scientific Reports}, 11\penalty0 (3520), 2021.
\newblock \doi{https://doi.org/10.1038/s41598-021-83119-x}.

\bibitem[Cabral-Passos et~al.(2024)Cabral-Passos, Galves, Garcia, and Vargas]{CabralPassos2024}
P.~R. Cabral-Passos, A.~Galves, J.~E. Garcia, and C.~D. Vargas.
\newblock Limit theorems for sequences of random trees.
\newblock \emph{Scientific Reports}, 14\penalty0 (8446), 2024.
\newblock \doi{https://doi.org/10.1038/s41598-024-58203-7}.

\bibitem[Hernández et~al.(2024)Hernández, Galves, García, Gubitoso, and Vargas]{Hernandez2024}
N~Hernández, A~Galves, J.~E. García, M.~D. Gubitoso, and C.~D. Vargas.
\newblock Probabilistic prediction and context tree identification in the goalkeeper game.
\newblock \emph{Scientific Report}, 14\penalty0 (15467), 2024.
\newblock \doi{https://doi.org/10.1038/s41598-024-66009-w}.

\bibitem[Dehaene et~al.(2015)Dehaene, Meyniel, Wacongne, Wang, and Pallier]{Dehaene2015}
S.~Dehaene, F.~Meyniel, C.~Wacongne, L.~Wang, and C.~Pallier.
\newblock The neural representation of sequences: From transition probabilities to algebraic patterns and linguistic trees.
\newblock \emph{Neuron}, 88\penalty0 (1):\penalty0 2--19, 2015.
\newblock ISSN 0896-6273.
\newblock \doi{10.1016/j.neuron.2015.09.019}.
\newblock URL \url{https://www.sciencedirect.com/science/article/pii/S089662731500776X}.

\bibitem[Planton et~al.(2021)Planton, van Kerkoerle, Abbih, Maheu, Meyniel, Sigman, et~al.]{Planton2021}
S.~Planton, T.~van Kerkoerle, L.~Abbih, M.~Maheu, F.~Meyniel, M.~Sigman, et~al.
\newblock A theory of memory for binary sequences: Evidence for a mental compression algorithm in humans.
\newblock \emph{PLoS Computational Biology}, 17\penalty0 (1):\penalty0 e1008598, 2021.
\newblock \doi{10.1371/journal.pcbi.1008598}.
\newblock URL \url{https://doi.org/10.1371/journal.pcbi.1008598}.

\bibitem[Al-Roumi et~al.(2023)Al-Roumi, Planton, Wang, and Dehaene]{AlRoumi2023}
F.~Al-Roumi, S.~Planton, L.~Wang, and S.~Dehaene.
\newblock Brain‑imaging evidence for compression of binary sound sequences in human memory.
\newblock \emph{eLife}, 12:\penalty0 84376, 2023.
\newblock \doi{10.7554/eLife.84376}.

\bibitem[Rissanen(1983{\natexlab{a}})]{Rissanen1983}
Jorma Rissanen.
\newblock A universal prior for integers and estimation by minimum description length.
\newblock \emph{Annals of Statistics}, 11\penalty0 (2):\penalty0 416--431, June 1983{\natexlab{a}}.
\newblock \doi{10.1214/aos/1176346150}.
\newblock URL \url{https://doi.org/10.1214/aos/1176346150}.

\bibitem[Rissanen(1983{\natexlab{b}})]{Rissanen1983b}
J.~Rissanen.
\newblock A universal data compression system.
\newblock \emph{IEEE Transactions on Information Theory}, 29:\penalty0 656--664, 1983{\natexlab{b}}.
\newblock \doi{https://doi.org/10.1109/TIT.1983.1056741}.

\bibitem[Cuesta‑Albertos et~al.(2007)Cuesta‑Albertos, el~Barrio, Fraiman, and Matrán]{CuestaAlbertos2007}
J.~A. Cuesta‑Albertos, E.~el~Barrio, R.~Fraiman, and C.~Matrán.
\newblock The random projection method in goodness of fit for functional data.
\newblock \emph{Computational Statistics and Data Analysis}, 51\penalty0 (10):\penalty0 4814--4831, 2007.
\newblock ISSN 0167-9473.
\newblock \doi{10.1016/j.csda.2006.09.007}.

\bibitem[Henderson(1953)]{Henderson1953}
C.~R. Henderson.
\newblock Estimation of variance and covariance components.
\newblock \emph{Biometrics}, 9\penalty0 (2):\penalty0 226--252, 1953.
\newblock \doi{https://doi.org/10.2307/3001853}.

\bibitem["Wikistat"(2016)]{Wikistat2016}
"Wikistat".
\newblock Modèle à effets aléatoires et modèle mixte, 2016.
\newblock URL \url{http://wikistat.fr/pdf/st-m-modmixt6-modmixt.pdf}.

\bibitem[Cuesta-Albertos et~al.(2006)Cuesta-Albertos, Fraiman, and Ransford]{CuestaAlberto2021}
J.~A. Cuesta-Albertos, R.~Fraiman, and T.~Ransford.
\newblock Random projections and goodness-of-fit tests in infinite-dimensional spaces.
\newblock \emph{Bull. Braz. Math. Soc. New Ser.}, 37:\penalty0 477--501, 2006.
\newblock \doi{https://doi.org/10.1007/s00574-006-0023-0}.

\bibitem[David et~al.(2009)David, Ferrari, Fraiman, and Sued]{Balding2009}
B.~David, P.~A. Ferrari, R.~Fraiman, and M~Sued.
\newblock Limit theorems for sequences of random trees.
\newblock \emph{TEST}, 18\penalty0 (2):\penalty0 302--315, 2009.
\newblock \doi{https://doi.org/10.1007/s11749-008-0092-z}.

\bibitem[Bogaerts et~al.(2020)Bogaerts, Richter, Landau, and Frost]{Bogaerts2020}
L.~Bogaerts, C.~G. Richter, A.~N. Landau, and R.~Frost.
\newblock Beta-band activity is a signature of statistical learning.
\newblock \emph{Journal of Neuroscience}, 40\penalty0 (39):\penalty0 7523--7530, Sep 2020.
\newblock \doi{10.1523/JNEUROSCI.0771-20.2020}.
\newblock URL \url{https://doi.org/10.1523/JNEUROSCI.0771-20.2020}.
\newblock Epub 2020 Aug 21. PMID: 32826312; PMCID: PMC7511193.

\bibitem[Widmann et~al.(2015)Widmann, Schröger, and Maess]{Widmann2015}
A.~Widmann, E.~Schröger, and B.~Maess.
\newblock Digital filter design for electrophysiological data--a practical approach.
\newblock \emph{Journal of Neuroscience Methods}, 250:\penalty0 34--46, Jul 2015.
\newblock \doi{10.1016/j.jneumeth.2014.08.002}.
\newblock URL \url{https://doi.org/10.1016/j.jneumeth.2014.08.002}.
\newblock Epub 2014 Aug 13. PMID: 25128257.

\bibitem[Monroy et~al.(2019)Monroy, Meyer, Schröer, Gerson, and Hunnius]{Monroy2019}
C.~D. Monroy, M.~Meyer, L.~Schröer, S.~A. Gerson, and S.~Hunnius.
\newblock The infant motor system predicts actions based on visual statistical learning.
\newblock \emph{NeuroImage}, 185:\penalty0 947--954, Jan 2019.
\newblock \doi{10.1016/j.neuroimage.2017.12.016}.
\newblock URL \url{https://doi.org/10.1016/j.neuroimage.2017.12.016}.
\newblock Epub 2017 Dec 7. PMID: 29225063.

\bibitem[Sáringer et~al.(2023)Sáringer, Fehér, Sáry, and Kaposvári]{Saringer2023}
S.~Sáringer, Á. Fehér, G.~Sáry, and P.~Kaposvári.
\newblock Gamma oscillations in visual statistical learning correlate with individual behavioral differences.
\newblock \emph{Frontiers in Behavioral Neuroscience}, 17:\penalty0 1285773, 2023.
\newblock \doi{10.3389/fnbeh.2023.1285773}.
\newblock URL \url{https://doi.org/10.3389/fnbeh.2023.1285773}.

\bibitem[Wang et~al.(2017)Wang, Shen, Tino, Welchman, and Kourtzi]{Wang2017}
R.~Wang, Y.~Shen, P.~Tino, A.~E. Welchman, and Z.~Kourtzi.
\newblock Learning predictive statistics from temporal sequences: Dynamics and strategies.
\newblock \emph{Journal of Vision}, 17\penalty0 (12):\penalty0 1--1, 2017.
\newblock \doi{10.1167/17.12.1}.
\newblock URL \url{https://doi.org/10.1167/17.12.1}.

\bibitem[Nissen and Bullemer(1987)]{Nissen1987}
M.~J. Nissen and P.~Bullemer.
\newblock Attentional requirements of learning: Evidence from performance measures.
\newblock \emph{Cognitive Psychology}, 19:\penalty0 1--32, 1987.
\newblock \doi{10.1016/0010-0285(87)90002-8}.
\newblock URL \url{https://doi.org/10.1016/0010-0285(87)90002-8}.

\bibitem[Hunt and Aslin(2001)]{Hunt2001}
R.~H. Hunt and R.~N. Aslin.
\newblock Statistical learning in a serial reaction time task: Access to separable statistical cues by individual learners.
\newblock \emph{Journal of Experimental Psychology: General}, 130:\penalty0 658--680, 2001.
\newblock \doi{10.1037/0096-3445.130.4.658}.
\newblock URL \url{https://doi.org/10.1037/0096-3445.130.4.658}.

\bibitem[Frost et~al.(2015)Frost, Armstrong, Siegelman, and Christiansen]{Frost2015}
R.~Frost, B.~C. Armstrong, N.~Siegelman, and M.~H. Christiansen.
\newblock Domain generality versus modality specificity: The paradox of statistical learning.
\newblock \emph{Trends in Cognitive Sciences}, 19:\penalty0 117--125, 2015.
\newblock \doi{10.1016/j.tics.2014.12.010}.
\newblock URL \url{https://doi.org/10.1016/j.tics.2014.12.010}.

\bibitem[Notebaert et~al.(2009)]{Notebaert2009}
W.~Notebaert et~al.
\newblock Post-error slowing: An orienting account.
\newblock \emph{Cognition}, 111:\penalty0 275--279, 2009.
\newblock \doi{10.1016/j.cognition.2009.01.008}.
\newblock URL \url{https://doi.org/10.1016/j.cognition.2009.01.008}.

\bibitem[Bestmann et~al.(2008)Bestmann, Harrison, Blankenburg, Mars, Haggard, Friston, and Rothwell]{Bestamann2008}
Sven Bestmann, Lee~M. Harrison, Felix Blankenburg, Rogier~B. Mars, Patrick Haggard, Karl~J. Friston, and John~C. Rothwell.
\newblock Influence of uncertainty and surprise on human corticospinal excitability during preparation for action.
\newblock \emph{Current Biology}, 18\penalty0 (10):\penalty0 775--780, 2008.
\newblock ISSN 0960-9822.
\newblock \doi{https://doi.org/10.1016/j.cub.2008.04.051}.
\newblock URL \url{https://www.sciencedirect.com/science/article/pii/S096098220800537X}.

\bibitem[Sznabel et~al.(2023)Sznabel, Land, Kopp, et~al.]{Sznabel2023}
D.~Sznabel, R.~Land, B.~Kopp, et~al.
\newblock The relation between implicit statistical learning and proactivity as revealed by eeg.
\newblock \emph{Scientific Reports}, 13:\penalty0 15787, 2023.
\newblock \doi{10.1038/s41598-023-42116-y}.
\newblock URL \url{https://doi.org/10.1038/s41598-023-42116-y}.

\bibitem[Galves and Löcherbach(2008)]{Galves2008}
A.~Galves and E.~Löcherbach.
\newblock Stochastic chains with memory of variable length.
\newblock \emph{TICSP Series}, 38:\penalty0 117--133, 2008.

\bibitem[Galves et~al.(2012)Galves, Galves, García, Garcia, and Leonardi]{Galves2012}
A.~Galves, C.~Galves, J.~E. García, N.~L. Garcia, and F.~Leonardi.
\newblock Context tree selection and linguistic rhythm retrieval from written texts.
\newblock \emph{The Annals of Applied Statistics}, 6\penalty0 (1):\penalty0 186--209, 2012.
\newblock \doi{10.1214/11-AOAS518}.
\newblock URL \url{https://doi.org/10.1214/11-AOAS518}.

\bibitem[Galves and Löcherbach(2013)]{Galves2013}
A.~Galves and E.~Löcherbach.
\newblock Infinite systems of interacting chains with memory of variable length—a stochastic model for biological neural nets.
\newblock \emph{Journal of Statistical Physics}, 151:\penalty0 896--921, 2013.
\newblock \doi{10.1007/s10955-013-0733-3}.
\newblock URL \url{https://doi.org/10.1007/s10955-013-0733-3}.

\bibitem[Garivier and Leonardi(2011)]{Garivier2011}
A.~Garivier and F.~Leonardi.
\newblock Context tree selection: A unifying view.
\newblock \emph{Stochastic Processes and their Applications}, 121:\penalty0 2488--2506, 2011.
\newblock \doi{10.1016/j.spa.2011.06.012}.
\newblock URL \url{https://doi.org/10.1016/j.spa.2011.06.012}.

\bibitem[Dawson(1947)]{Dawson1947}
C.~X. Dawson.
\newblock The averaging method for enhancing the event-related potential in electroencephalography.
\newblock \emph{Journal of Clinical Neurophysiology}, 4\penalty0 (2):\penalty0 117--125, 1947.

\bibitem[Benjamini and Yekutieli(2001)]{Benjamini2001}
Yoav Benjamini and Daniel Yekutieli.
\newblock The control of the false discovery rate in multiple testing under dependency.
\newblock \emph{The Annals of Statistics}, 29\penalty0 (4):\penalty0 1165 -- 1188, 2001.
\newblock \doi{10.1214/aos/1013699998}.
\newblock URL \url{https://doi.org/10.1214/aos/1013699998}.

\bibitem[Walter et~al.(1964)Walter, Cooper, Aldridge, McCallum, and Winter]{Walter1964}
W.~G. Walter, R.~Cooper, V.~J. Aldridge, W.~C. McCallum, and A.~L. Winter.
\newblock Contingent negative variation: An electric sign of sensori‐motor association and expectancy in the human brain.
\newblock \emph{Nature}, 203\penalty0 (4943):\penalty0 380--384, 1964.
\newblock \doi{10.1038/203380a0}.

\end{thebibliography}

\break

\section*{Supplementary Material}

\subsection*{Methods}

Given the specificities of the Goalkeeper game, such as the variable response time latency and the absence of a rest period between trials, some adaptations of the classical method for recovering the event-related potential from the EEG data\citep{Dawson1947} were necessary. Here, we describe the entire procedure, which is illustrated in Figure S1 for the ERPs aligned with the button press. Given the inter-trial variability of the response times, the signal length from the feedback to the current trial response time varies from trial to trial (Figure S1). Figure S1A represents the EEG segments recorded from a single electrode/goalkeeper across trials. The EEG signals were baseline corrected using the time interval of 200ms before the last feedback end. 
Signals of different electrodes have the same length for the same trial. Therefore, as illustrated in Figure S1B, a single distribution of lengths was obtained for each goalkeeper, independently of the electrode, and the value corresponding to the 75th percentile of the distribution was chosen as the length of the goalkeeper's event-related potential. Figure S1C shows that trials with a signal length larger than $75\%$ were trimmed on the left, while those with signals smaller than $75\%$ were flagged. Flagged intervals were not considered in the point-by-point average. As a result, the final ERP for the electrode represented in Figure S1D was based on an average with more points closer to the response time than closer to the previous feedback. 
To obtain the grand-average ERP, i.e., the mean ERP for an electrode across participants, the smaller length of the individual ERPs across the subjects was chosen. As to the verification of the significance of the grand-average ERP, the usual $t$-test associated with the \citet{Benjamini2001} correction for multiple comparisons was adopted.  

\subsection*{Results}

Figure S2 shows the Event-Related Potentials (ERPs) per electrode at the time interval between the end of the feedback and the next button press. For most electrodes, a significant negative deflection is verified in the time interval between -400 and 0 ms. This negative deflection resembles the classic contingent negative variation \citep{Walter1964}, a motor preparation marker traditionally associated with motor preparation in paradigms where a visual cue is employed to implement a choice-based motor decision.

\begin{table}[H]
\centering
\begin{tabular}{|l|l|l|l|l|l|}
\hline
Electrode & Intercept & Coefficient & $t$-statistics & $df$ & $p$-value              \\
\hline
Fp1       & 0.787     & -0.040      & -1.155       & 23 & 0.259                \\
Fp2       & 0.848     & -0.114      & -3.845       & 23 & $0.8 \times 10^{-3}$ \\
F7        & 0.836     & -0.100      & -3.355       & 23 & 0.002                \\
F3        & 0.826     & -0.084      & -1.939       & 23 & 0.064                \\
Fz        & 0.809     & -0.063      & -1.864       & 23 & 0.075                \\
F4        & 0.866     & -0.127      & -4.046       & 23 & $0.5 \times 10^{-3}$ \\
F8        & 0.807     & -0.068      & -1.250       & 23 & 0.223                \\
FT9       & 0.749     & 0.003       & 0.076        & 23 & 0.939                \\
Fc5       & 0.749     & 0.002       & 0.075        & 23 & 0.940                \\
Fc1       & 0.810     & -0.065      & -2.3493      & 23 & 0.027                \\
Fc2       & 0.801     & -0.051      & -1.699       & 23 & 0.102                \\
Fc6       & 0.791     & -0.044      & -0.968       & 23 & 0.342                \\
FT10      & 0.746     & 0.007       & 0.198        & 23 & 0.844                \\
T7        & 0.775     & -0.027      & -0.712       & 23 & 0.483                \\
C3        & 0.777     & -0.026      & -0.620       & 23 & 0.541                \\
Cz        & 0.831     & -0.090      & -2.913       & 23 & 0.007                \\
C4        & 0.828     & -0.079      & -2.748       & 23 & 0.011                \\
T8        & 0.798     & -0.055      & -1.155       & 23 & 0.259                \\
TP9       & 0.823     & -0.085      & -2.555       & 23 & 0.017                \\
Cp5       & 0.804     & -0.060      & -1.534       & 23 & 0.138                \\
Cp1       & 0.806     & -0.065      & -1.718       & 23 & 0.099                \\
Cp2       & 0.844     & -0.105      & -4.835       & 23 & $0.7 \times 10^{-4}$ \\
Cp6       & 0.725     & 0.029       & 0.657        & 23 & 0.517                \\
TP10      & 0.782     & -0.033      & -1.026       & 23 & 0.315                \\
P7        & 0.839     & -0.104      & -2.647       & 23 & 0.014                \\
P3        & 0.842     & -0.101      & -5.283       & 23 & $2.3 \times 10^{-5}$ \\
Pz        & 0.836     & -0.099      & -3.001       & 23 & 0.006                \\
P4        & 0.779     & -0.029      & -0.815       & 23 & 0.423                \\
P8        & 0.735     & 0.018       & 0.041        & 23 & 0.666                \\
O1        & 0.801     & -0.057      & -1.604       & 23 & 0.122                \\
Oz        & 0.843     & -0.099      & -4.029       & 23 & $0.5 \times 10^{-4}$  \\
O2        & 0.755     & -0.003      & -0.096       & 23 & 0.923           \\    
\hline
\end{tabular}
    {\caption*{ \textbf{Table TS1:} Mean intercept, angular coefficients, and statistics for the angular coefficients of the Linear Mixed Effects Model.}}
\end{table}

\begin{figure}[H]
    \centering
    \includegraphics[width=1\linewidth]{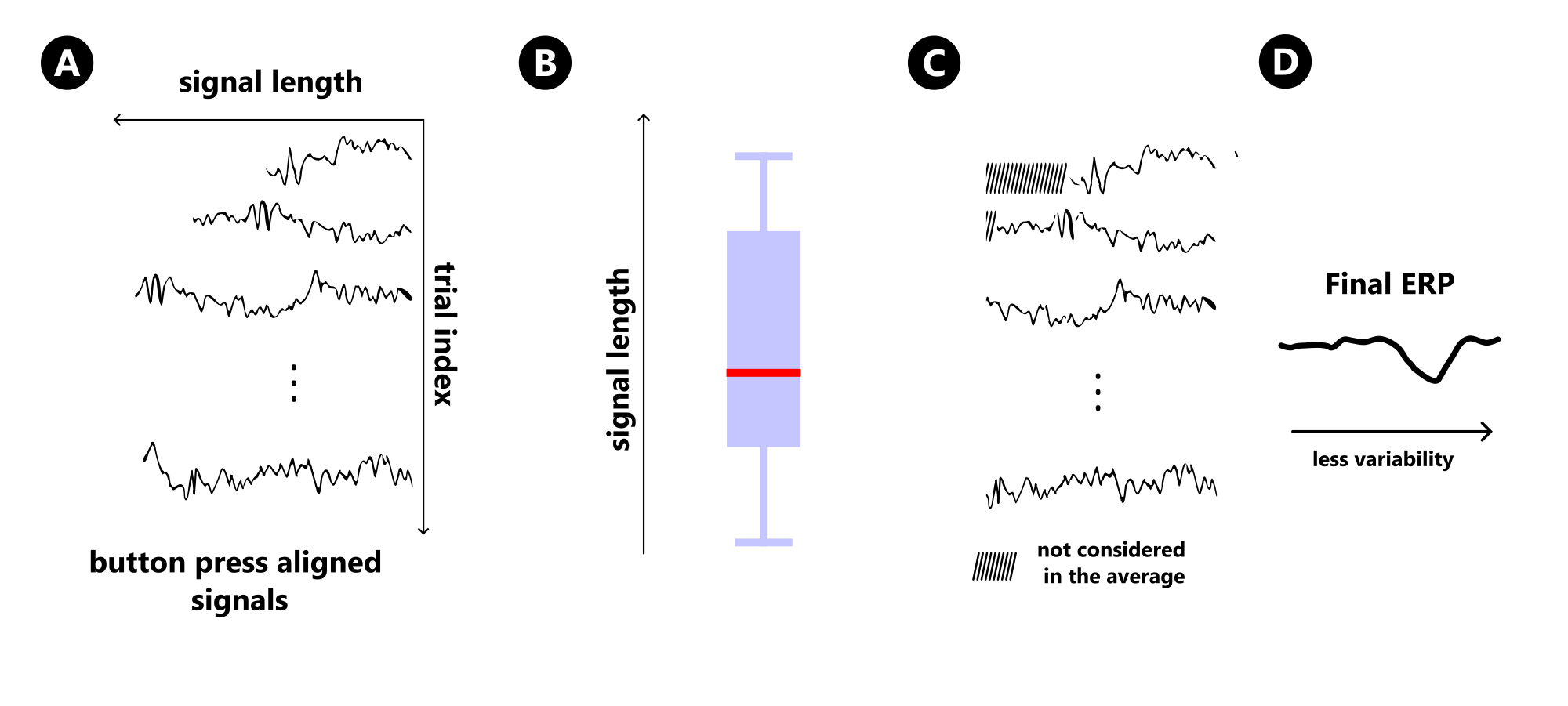}
    {\caption*{ \textbf{Figure S1: ERP estimation method for signals with different lengths.} A) Signals of a single electrode/goalkeeper at the time interval between the end of the feedback and the next button press (right alignment). The picture illustrates the different signal lengths given the variable response times across trials. B) To estimate the event-related potential (ERP) per goalkeeper, the $75\%$ percentile of the distribution of segment lengths was chosen as the ERPs final length. C) EEG signals with a length greater than the threshold were trimmed on the left, while signals with a length smaller than the threshold were flagged. D) The final ERP was calculated by using the traditional point-by-point average, but disregarding the flagged intervals of the smaller signals. As a consequence, the final ERP average has less variability closer to the button press }}
\end{figure}

\begin{figure}[H]
    \centering
    \includegraphics[width=1\linewidth]{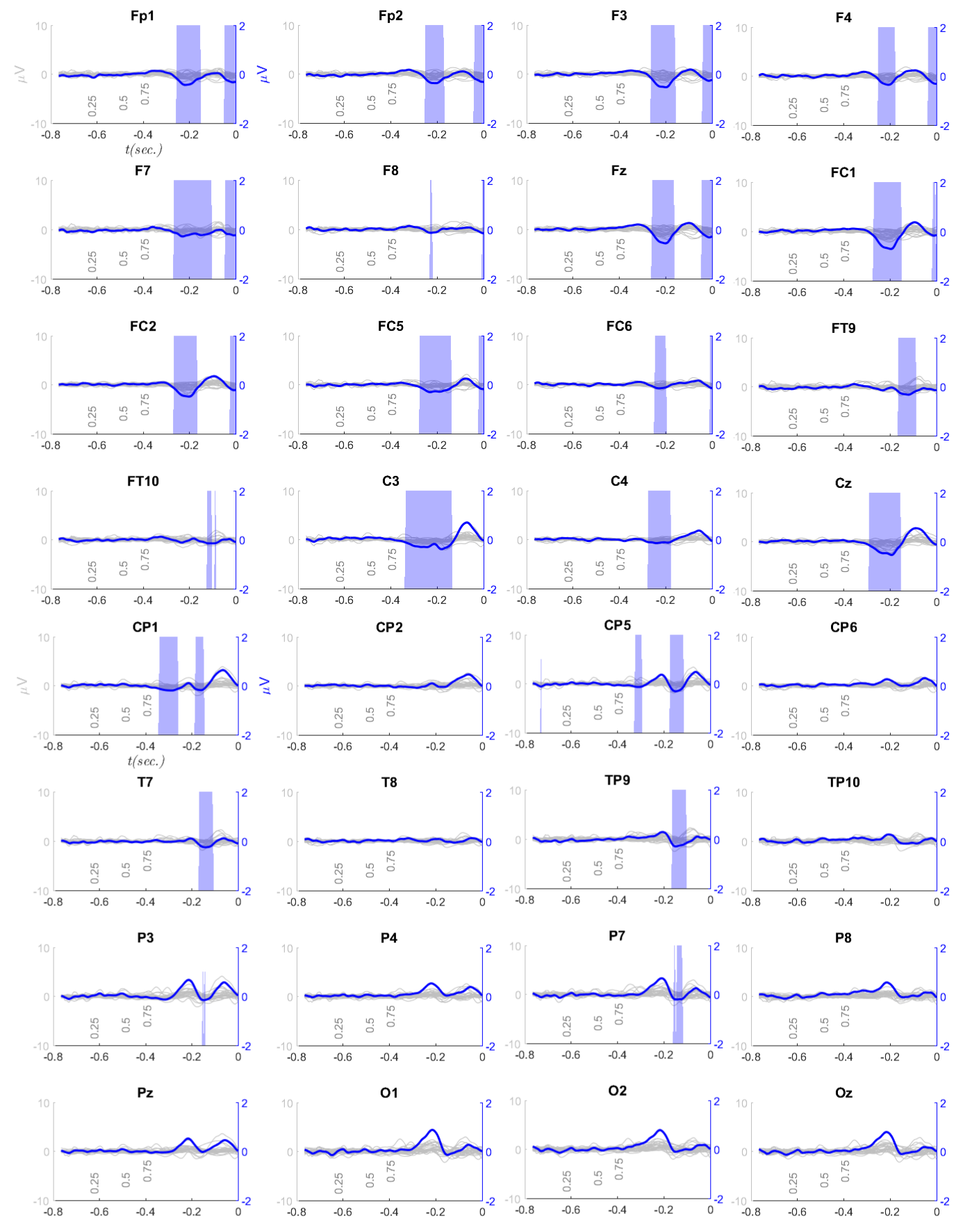}
    {\caption*{ \textbf{Figure S2: Event-Related Potentials (ERPs) at the time interval between the end of the feedback and the next button press.} Electrodes are identified by their respective labels in the 10-20 system. For each electrode, the left vertical axis shows the ERP amplitude per goalkeeper (gray). The right vertical axis shows the amplitude for the grand-average ERP (blue). Significant intervals ($p < 0.05$) of the grand-average ERPs are indicated by the blue shaded bars. Given the variability of the response times, the percentile labels 0.75, 0.50, and 0.25 indicate the percentage of participants with available signal (from the right to the left). For most electrodes, a significant negative deflection is verified in the time interval ranging between -400 and 0 ms. 
 
}}
\end{figure}

\begin{figure}[h]
    \centering
    \includegraphics[width=1\linewidth]{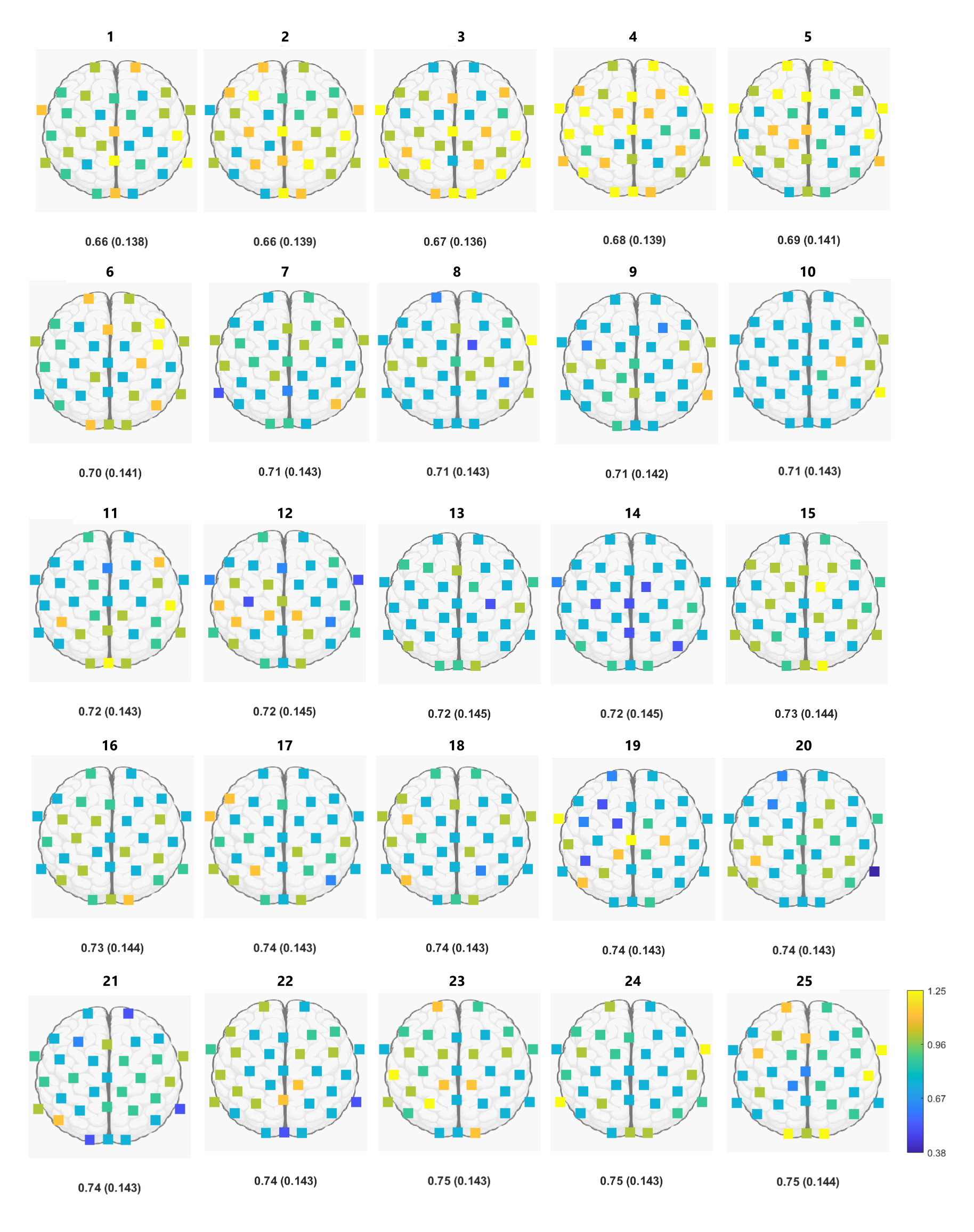}
    \caption*{\textbf{Figure S3: Time evolution of the Balding’s distance}. Context trees were estimated from sliding windows of 300 trials with a 50-trial displacement. Then, the distances between the penalty taker’s context tree and each goalkeeper’s estimated tree were calculated per window/electrode. Top numbers indicate the window’s index. Blue indicates distances closer to zero, while yellow indicates distances far from zero. The mean success rate is presented along with the standard deviation at the bottom of each topo-plot. The topo-plot time evolution suggests a reduction of the median distances for several electrodes along with an increase in the success rate.}
    \label{fig:S3}
\end{figure}

\begin{landscape}
\begin{figure}[htbp]
    \centering
    \includegraphics[width=1\linewidth]{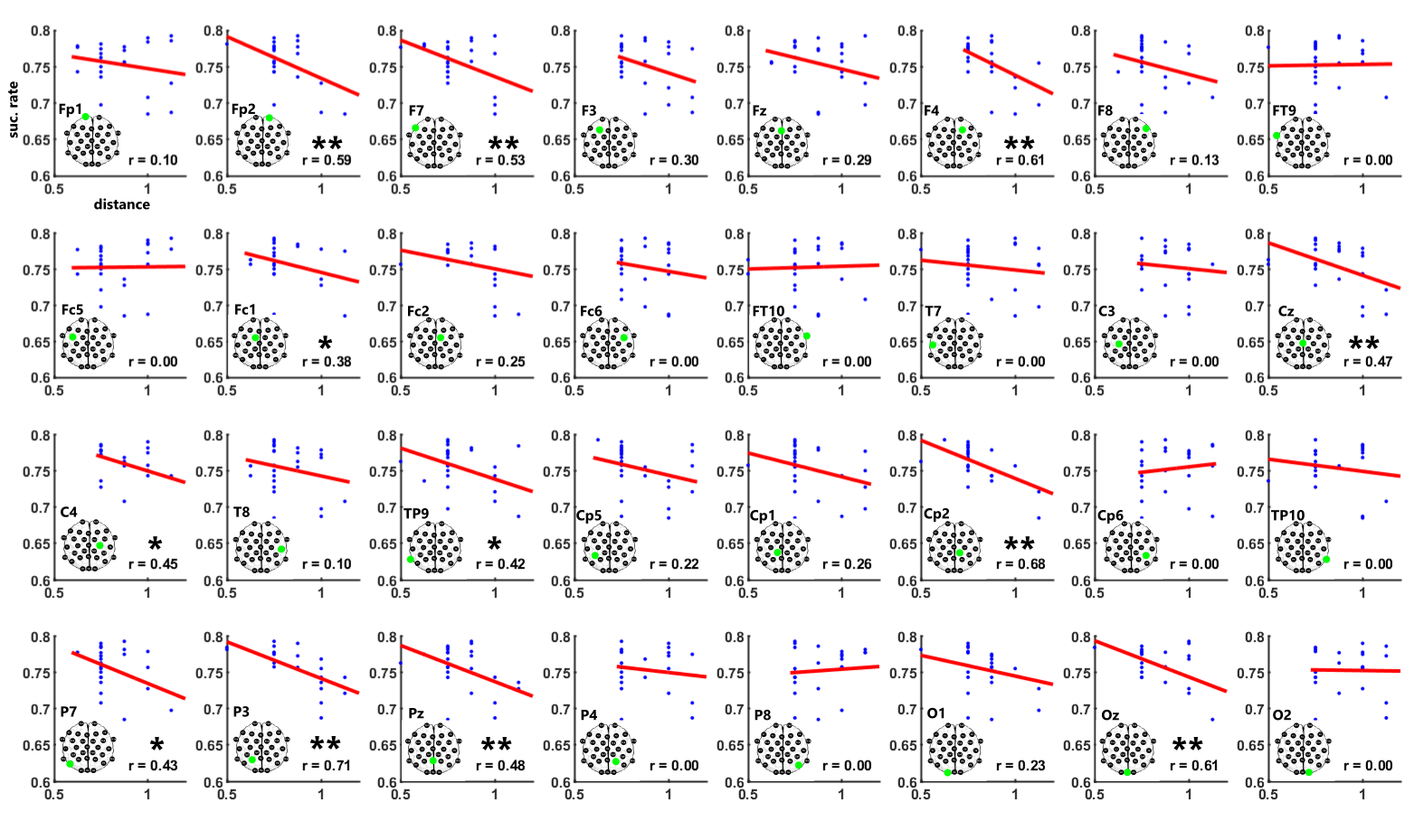}
    \caption*{\textbf{Figure S4: Success rate as a function of Balding’s distance.} A Linear Mixed Effects model was used for evaluating the predictability of the success rate based on Balding’s distance. Each graph corresponds to data of a single electrode highlighted in green in the bottom-left cartoon. In every graph, each blue dot represents a median-based centroid of the distances paired with the median-based success rate centroid per window. The significance of the correlations is represented as * for $p<0.05$ and ** for $p<0.01$. The absolute values of the correlations (r) are presented in the bottom right corner of each graph. An inverse correlation between success rates and the Balding distance for specific electrodes indicates that the corresponding brain regions are involved in the structural learning of the penalty taker's sequence.
}
    \label{fig:S4}
\end{figure}
\end{landscape}

\end{document}